\documentclass[12pt]{iopart}

\usepackage{iopams}
\usepackage{graphicx}
\newcommand{\rp}{\emph{r}-process}
\newcommand{\ri}{\emph{r}-I}
\newcommand{\rii}{\emph{r}-II}
\usepackage{xcolor,soul}

\begin{document}

\title[Characterizing \emph{r}-Process Sites]{Characterizing \emph{r}-Process Sites through Actinide Production}

\author{Erika M.\ Holmbeck$^{1,2}$, Rebecca Surman$^{1,2}$, Anna Frebel$^{3,2}$, G.\ C.\ McLaughlin$^{4,2}$, Matthew R.\ Mumpower$^{5,6,2}$, Trevor M.\ Sprouse$^1$, Toshihiko Kawano$^{5}$, Nicole Vassh$^{1,2}$, Timothy C.\ Beers$^{1,2}$}

\address{$^1$ Department of Physics, University of Notre Dame, Notre Dame, IN 46556, USA}
\address{$^2$ JINA Center for the Evolution of the Elements, USA}
\address{$^3$ Department of Physics and Kavli Institute for Astrophysics and Space Research, Massachusetts Institute of Technology, Cambridge, MA 02139, USA}
\address{$^4$ Department of Physics, North Carolina State University, Raleigh, NC 27695, USA}
\address{$^5$ Theoretical Division, Los Alamos National Laboratory, Los Alamos, NM, 87545, USA}
\address{$^6$ Center for Theoretical Astrophysics, Los Alamos National Laboratory, Los Alamos, NM, 87545, USA}

\ead{eholmbec@nd.edu}
\vspace{10pt}
\begin{indented}
\item[]November 2019
\end{indented}

\begin{abstract}
Of the variations in the elemental abundance patterns of stars enhanced with \rp\ elements, the variation in the relative actinide-to-lanthanide ratio is among the most significant.
We investigate the source of these actinide differences in order to determine whether these variations are due to natural differences in astrophysical sites, or due to the uncertain nuclear properties that are accessed in \rp\ sites.
We find that variations between relative stellar actinide abundances is most likely astrophysical in nature, owing to how neutron-rich the ejecta from an \rp\ event may be.
Furthermore, if an \rp\ site is capable of generating variations in the neutron-richness of its ejected material, then only one type of \rp\ site is needed to explain all levels of observed relative actinide enhancements.
\end{abstract}

%\maketitle

% ==========================================================
\section{Introduction}

The rapid neutron-capture (``\emph{r}") process is the physical mechanism by which about half the elements heavier than iron in the Solar System were created.
The \rp\ was first identified by \cite{b2fh} and \cite{cameron1957}, but it is still unclear where the \rp\ may occur astrophysically.
Outside of the Solar System, traces of the \rp\ lie in very metal-poor ($[{\rm Fe/H}]\footnote{$[{\rm A/B}] = \log(N_A/N_B)_* - \log(N_A/N_B)_\odot$, where $N$ is the number density of an element in the star (*) compared to the Sun ($\odot$).}=-2$) stars that show an enhancement of the heavy, \rp\ elements in their photospheres.
Stars with low metallicity indicate few nucleosynthetic events by supernovae that would otherwise release an abundance of iron into the primordial gas of the interstellar medium.
In this way, the elemental abundances in low-metallicity stars record the chemical signatures of nucleosynthetic events preceding their formation.

Of metal-poor stars, less than about 20\% show an enhancement of \rp\ elements in their photospheres.
The level of \rp\ enhancement in a star is quantified by the europium ($_{63}$Eu) abundance since Eu in the Solar System was predominantly made by the \rp, and it is one of the easiest elements to observe with high-resolution spectroscopy using ground-based telescopes.
The relative level of Eu to Fe abundance in a star is a proxy for how much \rp\ enhancement preceded the star's formation compared to the chemical evolution of that gas from supernova events.
Stars over-enhanced with Eu relative to Fe compared to the Sun are called ``\rp\ enhanced" and are divided into two categories: ``\ri,'' with $0.3 < [{\rm Eu/Fe}] \leq 1.0$ (i.e., between a factor of 2 and 10 greater than the Solar value), and ``\rii,'' with $[{\rm Eu/Fe}] > 1.0$ (i.e., over a factor of 10 greater than the Solar System).
These \ri\ and \rii\ stars are relics of prolific \rp\ event(s) that occurred before the gas was enriched by supernovae, and are therefore considered tracers of nearly pure \rp\ events.

\begin{figure}[t]
	\centering
	\includegraphics[width=0.7\textwidth]{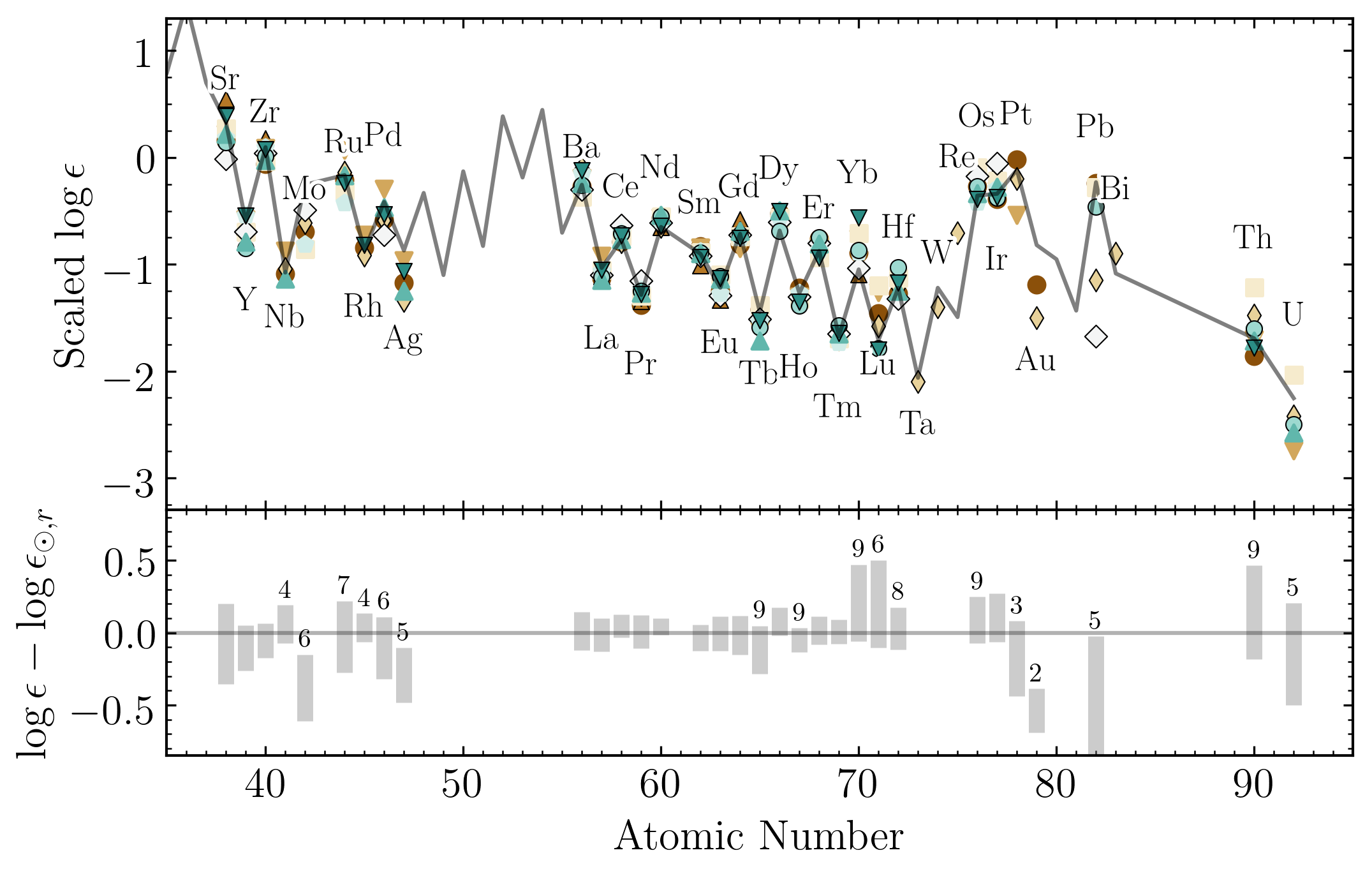}
	\caption{Abundance patterns of 10 \rii\ stars (CS22892-052 \cite{sneden2008}, CS22953-003 \cite{roederer2014b}, CS29497-004 \cite{hill2017}, CS31082-001 \cite{siqueira2013}, J0954+5246 \cite{holmbeck2018}, HE2327-5642 \cite{mashonkina2010}, HE2252-4225 \cite{mashonkina2014}, J2038$-$0023 \cite{placco2017}, HE1523-0901 \cite{frebel2007}, and J1538$-$1804 \cite{sakari2018}) scaled to their average difference from the Solar value between Eu and Lu (top), and the range of abundances for each element compared to the Solar System abundance (bottom). Numbers above the ranges in the bottom panel indicate how many stars from the upper panel have a measurement of that element. When not indicated, all 10 stars have a measured abundance.\label{fig:rii_patterns}}
	\end{figure}

Scaled to Eu, the relative abundance patterns of the \ri\ and \rii\ stars are strikingly similar between the second and third \rp\ peaks at $(Z,A)\sim (54,130)$ and $(78,195)$, respectively.
However, large variations exist in the actinide elements ($_{90}$Th and $_{92}$U) and the light \rp\ elements in the Sr-Y-Zr group ($Z=38$--40).
Figure~\ref{fig:rii_patterns} shows the scaled abundance patterns of several \rii\ stars as well as the variation across the stars for each element.
Stars with an over-abundance of [Th/Eu] relative to the Solar System are considered ``actinide-boost'' stars, occurring in about 30\% of \rp\ enhanced stars.

% ==========================================================
\subsection{Cosmochronometry}

Since the observable actinides, $^{232}$Th and $^{238}$U, have very long halflives (14.0~Gyr and 4.47~Gyr, respectively), radioactive decay principles can be applied to approximate an age for the material in stars in which both of these elements are observed.
Compared to some stable element co-produced by the \rp\ (typically Eu), the radioactive decay ages can be calculated in three ways:
	\begin{eqnarray}
	t = 46.67~\textrm{Gyr} \left[\log\epsilon\left(\textrm{Th/Eu}\right)_0 - \log\epsilon\left(\textrm{Th/Eu}\right)_{\textrm{\scriptsize obs}}\right] \label{eqn:theu}\\
	t = 14.84~\textrm{Gyr} \left[\log\epsilon\left(\textrm{U/Eu}\right)_0 - \log\epsilon\left(\textrm{U/Eu}\right)_{\textrm{\scriptsize obs}}\right] \label{eqn:ueu}\\
	t = 21.80~\textrm{Gyr} \left[\log\epsilon\left(\textrm{U/Th}\right)_0 - \log\epsilon\left(\textrm{U/Th}\right)_{\textrm{\scriptsize obs}}\right] \label{eqn:uth},
	\end{eqnarray}
where $\log\epsilon\left(\textrm{X/Eu}\right)_0$ is the initial production ratio corresponding to the formation of europium and element X by the \rp\ at $t=0$, and $\log\epsilon\left(\textrm{X/Eu}\right)_{\textrm{\scriptsize obs}}$ is the ratio derived from observations (present day).
By determining the observed ratio from stellar spectra, and assuming some initial production ratio, the time $t$ for which the radioactive element X has decayed can be calculated.
If the initial production ratios are a good descriptor of the progenitor \rp\ site, then all three of these ages should agree to the same value and be within physical limits (i.e., older than 0~Gyr, but no older than the age of the Universe).
Historically, production ratios are used from \rp\ waiting point calculations or \rp\ simulations of supernovae \cite{wanajo2002,farouqi2010}.
For most \rp\ enhanced stars, ages calculated with these production ratios are relatively consistent.
However, the same calculation with actinide-boost stars yields extremely different ages.
Notably, the Th/Eu age (Equation~\ref{eqn:theu}) is often negative, unphysically implying that actinide-boost stars are yet to be created.
Nevertheless, the U/Th age (Equation~\ref{eqn:uth}) consistently produces accurate ages for both actinide-boost and non-actinide-boost stars alike.

% ==========================================================
\subsection{\emph{r}-Process Sites}

The recent gravitational wave event GW170817 proved that neutron-star mergers (NSMs) occur in the Universe \cite{abbott2017}.
Follow-up on the electromagnetic (``kilonova") component to this event revealed evidence that the lanthanide elements were created in the merger, confirming that NSMs are one \rp\ site \cite{cowperthwaite2017,drout2017,kasen2017}.
Although lanthanide production was deduced from the event, the extent of actinide production (if any) by NSMs is unknown.
In this study, we find initial production ratios from NSM sites and apply those production ratios to investigate whether NSMs may be the source of the actinide-boost phenomenon observed in some \rp\ enhanced stars.

% ==========================================================
\section{Nuclear Physics Uncertainties}
\label{sec:nuc}

	\begin{figure}[t]
	\centering
	\includegraphics[width=0.65\textwidth]{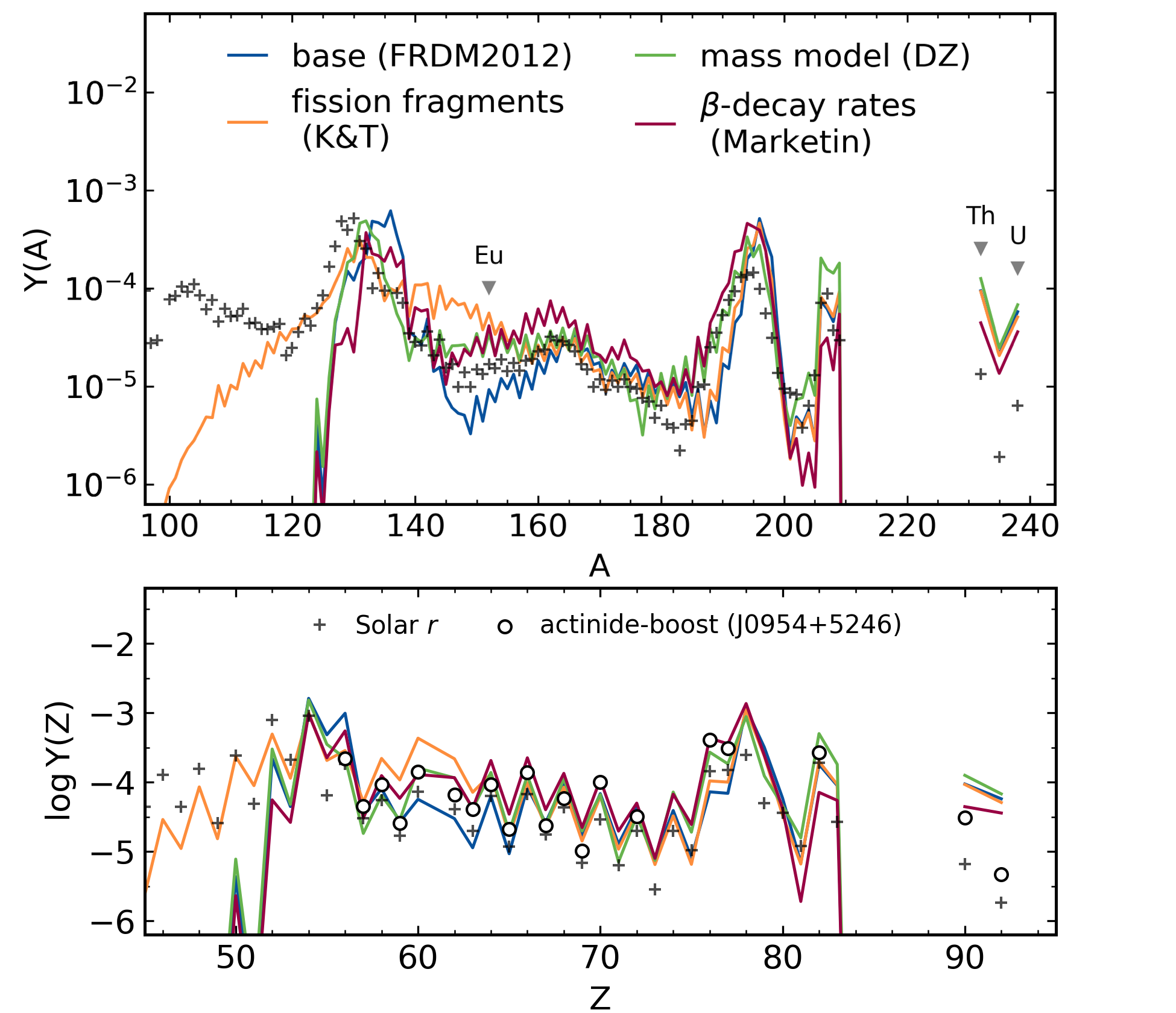}
	\caption{Final isotopic (top) and elemental (bottom) abundance patterns using the four variations on nuclear input described in Section~\ref{sec:nuc}. For guidance, the locations of the Eu, Th, and U isotopes are indicated. The scaled abundances for the actinide-boost star J0954$+$5246 are shown in the lower panel for comparison.\label{fig:nuc}}
	\end{figure}

In this study, we run nucleosynthesis simulations using the network code Portable Routines for Integrated nucleoSynthesis Modeling (PRISM, \cite{holmbeck2019a} and references therein).
We use a trajectory from the NSM simulations of S.\ Rosswog \cite{rosswog2013,piran2013} with recommended electron fraction of $Y_e = 0.035$.
For nuclear reaction and decay data, we start with JINA Reaclib nuclear reaction database \cite{cyburt2010}, adding relevant \rp\ nuclear data calculated as self-consistently as possible.
We calculate neutron-capture and neutron-induced fission rates self-consistently from input nuclear masses using the Los Alamos National Laboratory statistical Hauser-Feshbach code \cite{kawano2016}.
Where available, known masses and decay rates are used from the Atomic Mass Evaluation and Nubase2016 \cite{audi2017}. 
(For details, see \cite{holmbeck2019a}.)
We test four different variations on the nuclear physics input to test the sensitivity of actinide production on nuclear data:
	\begin{enumerate}
	\item Baseline: FRDM2012 masses, with $\beta$-decay rates from \cite{moller2019}, and a simple, symmetric description for fission fragments for all fission channels.
	\item DZ: As in the baseline case, but change nuclear masses according to the Duflo-Zuker mass model \cite{duflo1995}. Reaction and decay rates are recalculated using these masses, with $\beta$-strength functions and fission barrier heights kept the same as in the baseline case.
	\item K\&T: As in the baseline case, but change the fission fragment distribution to the double-Gaussian distributions described by \cite{kodama1975}.
	\item Marketin: As in the baseline case, but change the $\beta$-decay rates from \cite{moller2019} to those of \cite{marketin2016}.
	\end{enumerate}

Figure~\ref{fig:nuc} shows the final (1~Gyr) isotopic and elemental abundances for each of these four cases.
The elemental patterns are also compared to the abundances of an extremely actinide-boost star, J0954$+$5246 \cite{holmbeck2018}.
In every case, the actinides are overproduced compared to the observations, and the production of Eu varies dramatically.
Next, we test if astrophysical variations rather than differences in the assumed nuclear properties of nuclei far from stability can produce abundances in agreement with actinide-boost stars.

% ==========================================================
\section{Astrophysical Variations}
\label{sec:ye_variations}

	\begin{figure}[t]
	\centering
	\includegraphics[height=0.24\textheight]{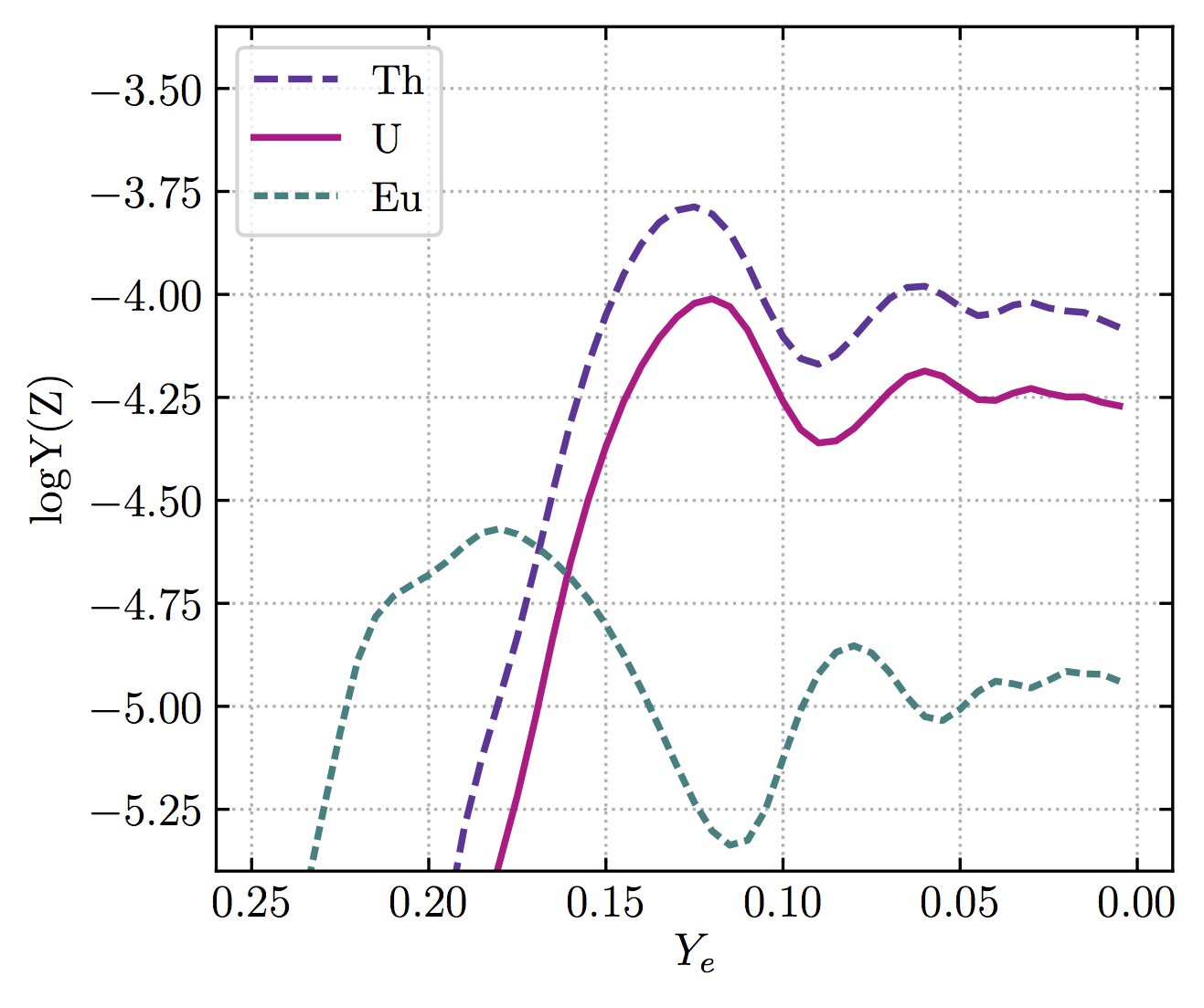}
	\includegraphics[height=0.24\textheight]{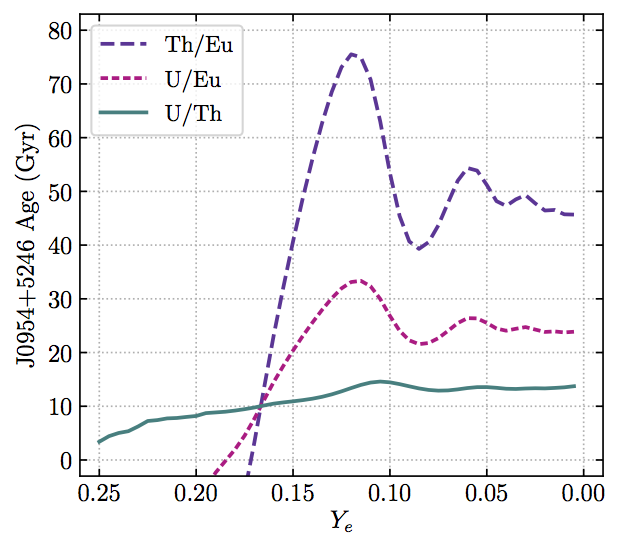}
	\caption{Left: final Th, U, and Eu abundances as a function of initial $Y_e$. Right: ages of J0954$+$5246 calculated using the corresponding Th/Eu, U/Eu, and U/Th production ratios at each $Y_e$. Note that only one $Y_e$ value gives consistent ages.\label{fig:ages_ye}}
	\end{figure}

The initial electron fraction, $Y_e$, taking any value between 0 and 1, describes the initial composition of the nuclear material involved in the \rp, with a lower $Y_e$ meaning more neutron-rich, and a higher $Y_e$ more proton-rich.
In this section, we change the initial nuclear composition by varying the $Y_e$ to investigate the sensitivity of actinide and lanthanide production on this astrophysical parameter.
For each case of nuclear input, we run 50 nucleosynthesis network simulations, changing the initial $Y_e$ from 0.005 to 0.250 in steps of 0.005.
Next, we take the final abundances from each of these simulations and use them as the initial production ratios to calculate ages for the actinide-boost star, J0954$+$5246.
Figure~\ref{fig:ages_ye} shows the final abundances and stellar ages calculated at each initial $Y_e$.
Only one value of $Y_e$ produces consistent ages, agreeing on an age of 11~Gyr (10~Gyr plus the 1~Gyr end-time of the nucleosynthesis simulations).
This particular $Y_e$ is at about 0.17, much greater than the suggested value of 0.035 from the hydrodynamical simulation of the dynamical ejecta.

% ==========================================================
\section{Actinide Dilution}

Rather than finding one $Y_e$ and one particular set of nuclear input needed to explain the actinide-boost, we consider a combination of $Y_e$ in a method we call ``Actinide-Dilution" (AD).
In the AD model, we choose a double-Gaussian distribution of $Y_e$, where one Gaussian represents the tidal/dynamical ejecta of an NSM, and the other the mass ejected by the disk wind.
Fits for the Gaussians are based on the models of \cite{bovard2017} using SFHO for the dynamical ejecta and the H000 model of \cite{lippuner2017} for the disk wind.
Then, the amplitudes of the Gaussians are adjusted such that the mass ratio between the dynamical and wind ejecta components is $m_{\textrm{\scriptsize w}}/m_{\textrm{\scriptsize dyn}} = 3$, which is what some studies suggest as the corresponding ejecta ratio from GW170817 \cite{rosswog2017,tanaka2017}.
In summary, this method generates a double-Gaussian approximation for the distribution of NSM ejecta as a function of $Y_e$.

	\begin{figure}[t]
	\centering
	\includegraphics[width=0.65\textwidth]{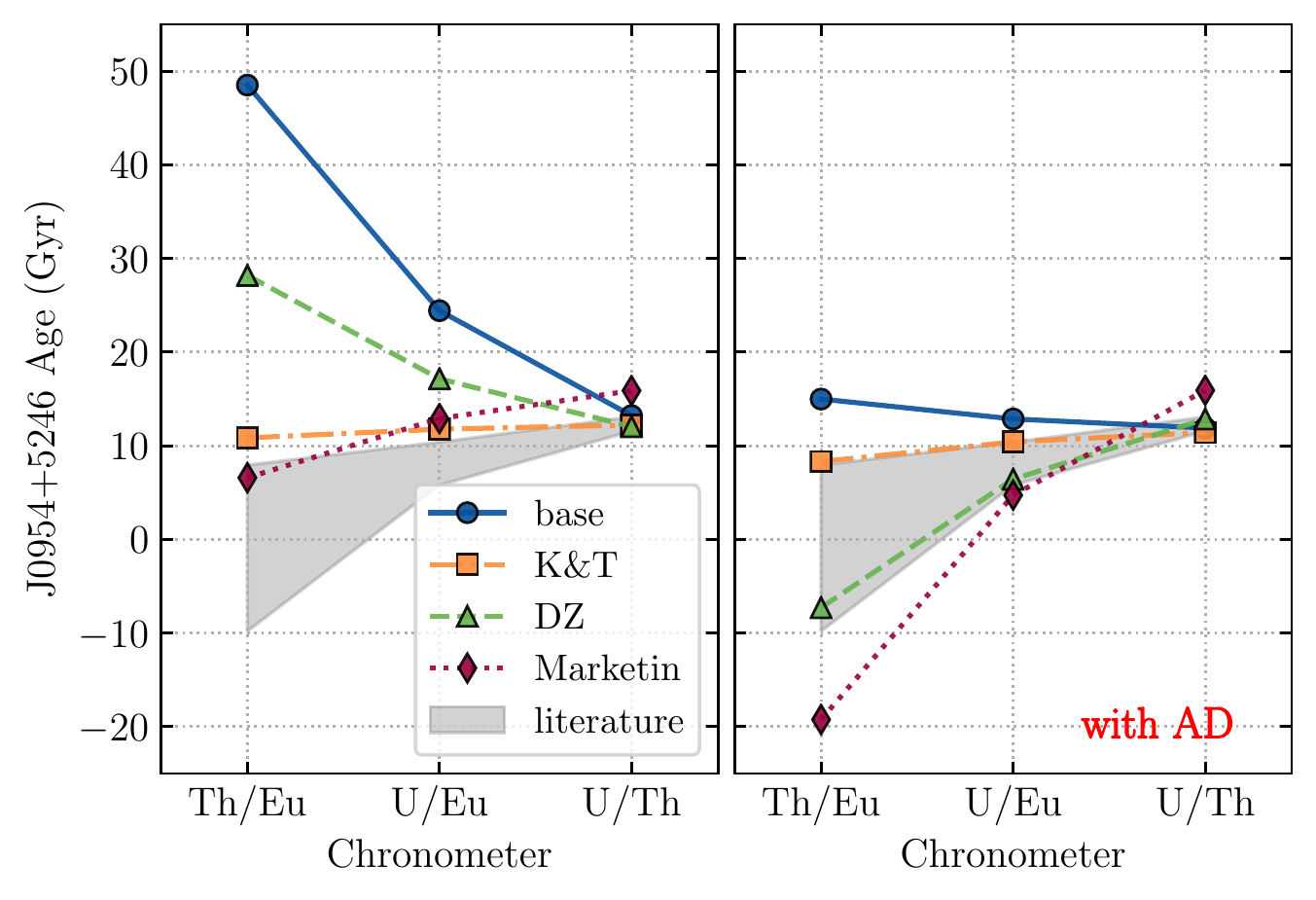}
	\caption{Left: ages for the actinide-boost star J0954$+$5246 calculated using Equations~\ref{eqn:theu}--\ref{eqn:uth} at a $Y_e$ of 0.035 for four choices of nuclear input. Right: the same ages, this time considering a distribution of $Y_e$ described by the AD model.\label{fig:ages_ad}}
	\end{figure}
	
Since the actinides are unstable, applying Equations~\ref{eqn:theu}--\ref{eqn:uth} will reveal if the production ratios from the nuclear variations (i.e., using the abundances from Figure~\ref{fig:nuc}) are a reasonable descriptor of the \rp\ that produced the material in actinide-boost stars.
The left panel of Figure~\ref{fig:ages_ad} shows the three ages calculated from Equations~\ref{eqn:theu}--\ref{eqn:uth} using the final abundances for each choice of nuclear physics input.
Recall that for the simulation to accurately describe the \rp\ event responsible for the actinide-boost star, all three ages must agree.
Using just the $Y_e=0.035$ abundances, only one set of nuclear data generates consistent ages: the K\&T case.
However, in this case, the broad fission fragment distribution is a poor match to the rest of the stellar abundance pattern of the actinide-boost star J0954$+$5246, and the Th/Eu and U/Eu are only consistent due to the artificially enhanced Eu abundances.
Using $\beta$-decay rates from Marketin produces the next best set of consistent ages.
However, the U/Th age (which is well-constrained by observations and other theoretical studies of production ratios) leads to an age of about 17~Gyr---much greater than the age of the Universe.

The right panel of Figure~\ref{fig:ages_ad} shows the ages calculated from the combination of $Y_e$ according to the AD model.
After applying the AD distribution of $Y_e$ (instead of a single $Y_e$) to each of the nuclear variations considered, we find that the actinide abundance and relative actinide-to-lanthanide ratio are lowered to levels that are roughly in agreement with abundances derived from observations of metal-poor stars.
The DZ mass model fared better than the FRDM2012 model in the single-$Y_e$ case, but leads to negative ages with the AD model.
Using the $\beta$-decay rates of Marketin also leads to negative Th/Eu ages and the anomalously high U/Th is not fixed by a combination of $Y_e$.
Stellar ages calculated from production ratios using the FRDM2012 mass model (the baseline case) generally lead to the most realistic and consistent ages.
Next, we take the AD model further by finding a description for the mass distribution as a function of $Y_e$ that fits different actinide variations.

% ==========================================================
\section{Actinide Dilution with Matching}

%	\begin{figure}[t]
%	\centering
%	\includegraphics[width=0.65\textwidth]{/Volumes/Files/Code/slides/img/group_patterns_GroupF_wes00.png}
%	\caption{Elemental abundance patterns showing actinide variation.\label{fig:patterns}}
%	\end{figure}

In this section, we extend the Actinide-Dilution model to more explicitly match an input abundance pattern in a new method we call the ``Actinide-Dilution with Matching" (ADM) model.
As a requirement for the model, the star must have a Th abundance determined, as well as Dy and Zr, representing the lanthanides and first \rp\ peak, respectively.
We use the abundances from the same actinide-boost star as above.

Three constraints are supplied to the ADM model: the relative Zr/Dy, Th/Dy, and U/Th abundances.
The Zr/Dy and Th/Dy abundances ratios are taken from the reported values of the actinide-boost star.
Not all stars have a reported measurement for U, so we choose the production ratio of $\log\epsilon(\rm{U/Th})=-0.25$ for this constraint.
Next, we run a series of simulations using PRISM and vary the initial $Y_e$ between 0.005 and 0.450, similar to the calculations in Section~\ref{sec:ye_variations}, this time instead using a trajectory consistent with the disk wind from a merger remnant instead of a dynamical trajectory (see \cite{holmbeck2019b} for details).
We also vary the input mass model, repeating all simulations with the Duflo-Zuker and the Hartree-Fock-Bogoliubov (HFB) \cite{hfb17} models after self-consistently recalculating all relevant nuclear data as described in Section~\ref{sec:nuc}.
Now, we have a set of final abundance patterns calculated across a range of $Y_e$ values for three different nuclear mass models.

The ADM model randomly selects 15 values of $Y_e$, combines the final associated abundances, then tests if the Zr/Dy, Th/Dy, and U/Th ratios agree with the input constraints from observationally derived abundances.
If all three are within about 0.2~dex of the input constraints, the set of $Y_e$ is kept and added to a cumulative mass distribution.
This set is then one realization of the ejecta from the event that could have produced the input abundances.
Otherwise, the model resamples a different set of 15 and tries again.
The model stops after it has accumulated 100 successes, totaling 1500 summed abundance patterns and a full distribution of $Y_e$.

	\begin{figure}[t]
	\centering
	\includegraphics[height=0.236\textheight]{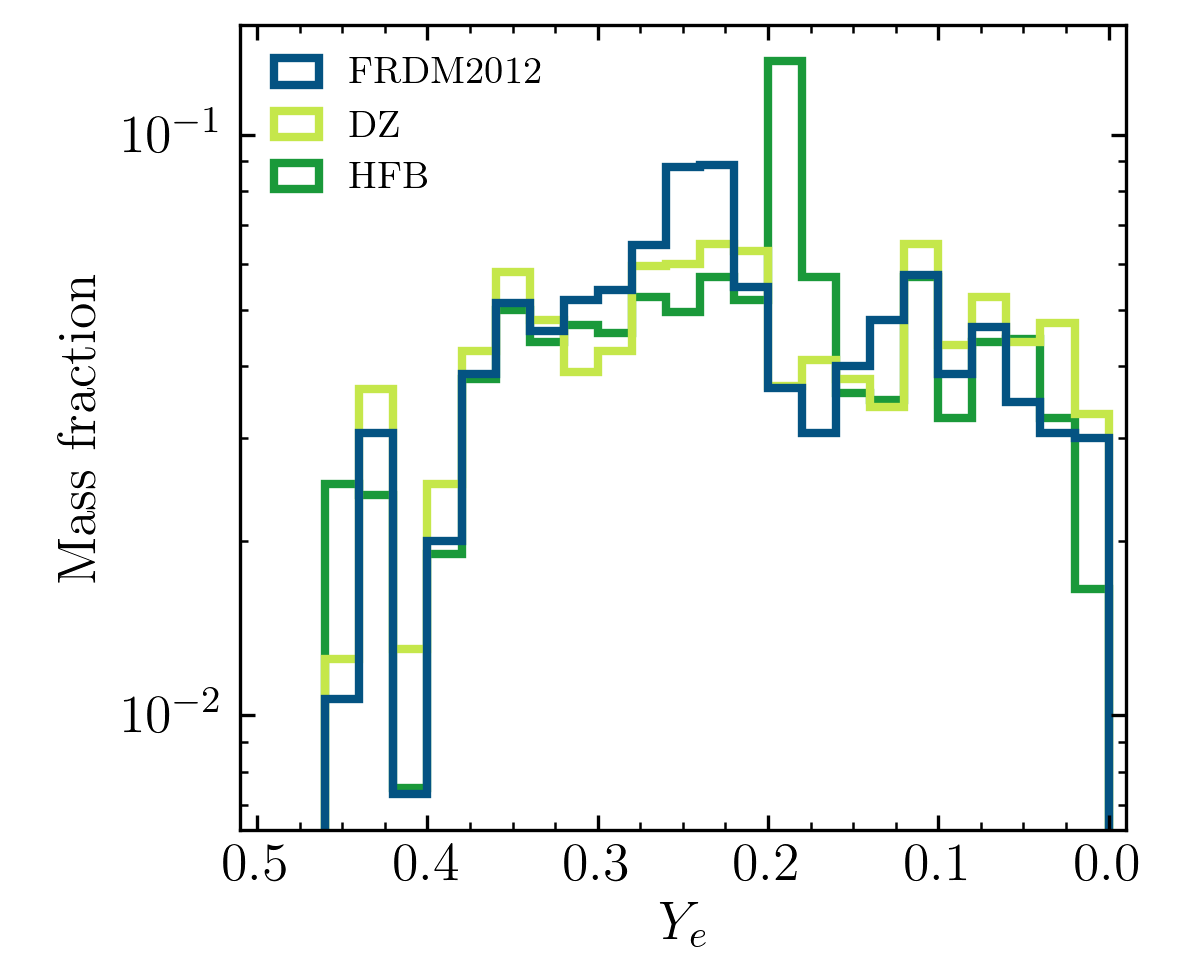}
	\includegraphics[height=0.24\textheight]{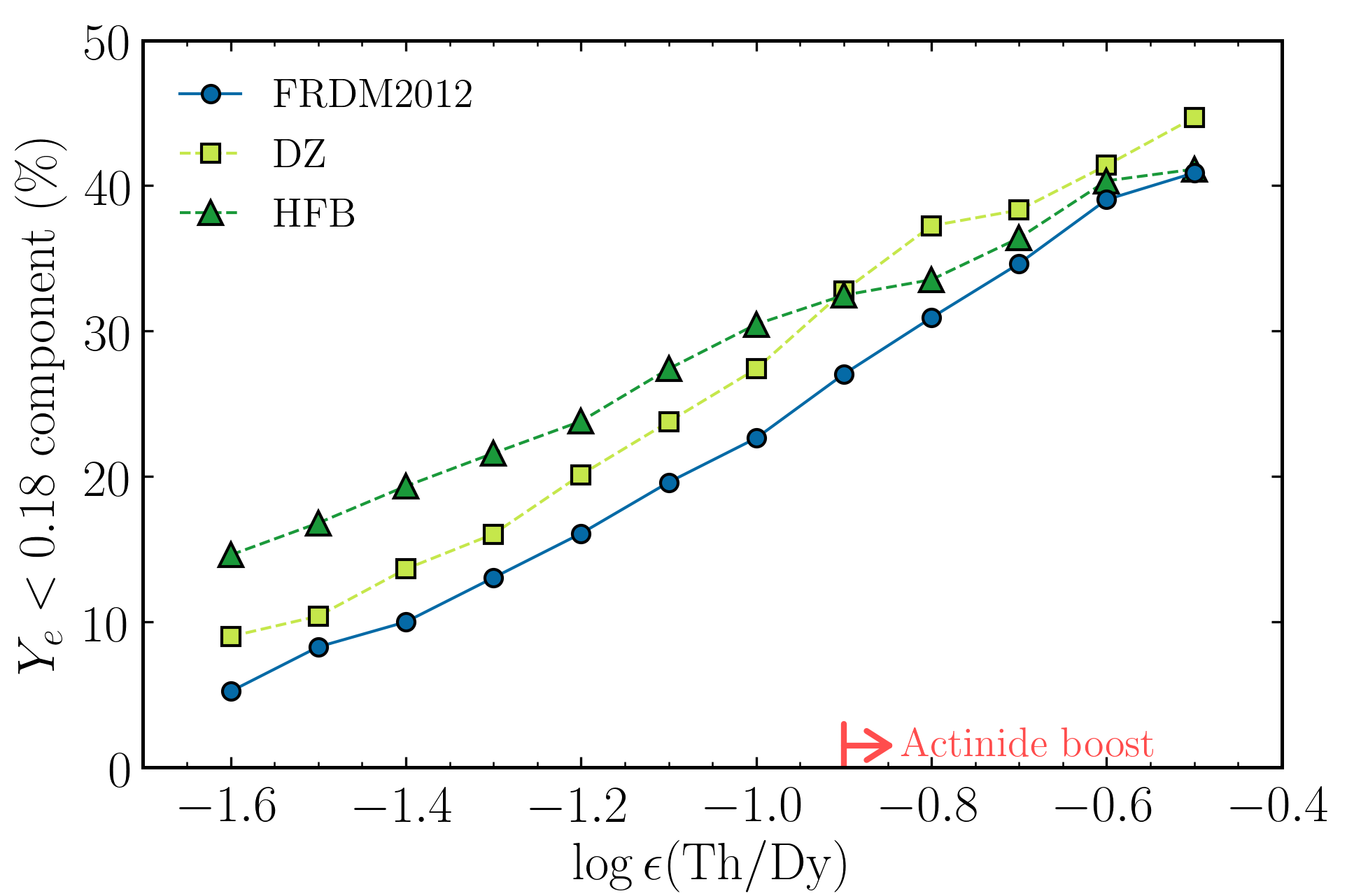}
	\caption{Left: mass fractions of \rp\ ejecta predicted by the ADM model using abundances from J0954$+$5246 as matching constrains with three different nuclear mass models. Right: the percentage of very low-$Y_e$ mass that may be ejected from an \rp\ site in order to produce the specified actinide-to-lanthanide abundance ratio ($\log\epsilon(\rm{Th/Dy})$).\label{fig:adm}}
	\end{figure}

The left panel of Figure~\ref{fig:adm} shows the mass distribution results for each mass model applied to the abundances for J0954$+$5246.
This mass distribution describes how the ejecta from an \rp\ event could have been distributed in order to match the input abundances.
We also test other input actinide values, rerunning the ADM model using a range of $\log\epsilon(\rm{Th/Dy})$ (actinide-to-lanthanide) ratios that spans over all current observationally derived $\log\epsilon(\rm{Th/Dy})$ abundances, including actinide-poor and actinide-boost.
Then, we look at the total contribution from the very neutron-rich component ($Y_e<0.18$).
The right panel of Figure~\ref{fig:adm} shows the percentage of the total ejecta mass that begins the \rp\ at very neutron-rich values for certain input relative actinide abundances.
The smoothly increasing trend for the low-$Y_e$ component as a function of actinide-to-lanthanide ratio is similar for each mass model, with about a 10\% maximum difference between the three models.
In the HFB case, the actinides are not produced as abundantly as in the FRDM2012 or DZ cases, and therefore more low-$Y_e$ material is allowed to contribute to the total, final actinide abundance.
However, the difference between an actinide-poor case ($\log\epsilon(\rm{Th/Dy})<-1.20)$ and an actinide-boost case ($\log\epsilon(\rm{Th/Dy}) > -0.90$) is only about 15\%.
In other words, an \rp\ event must eject roughly 15\% more of its material at very low values of $Y_e$ in order to change its produced abundances from actinide-poor to actinide-rich.

% ==========================================================
\section{Conclusions}

We have investigated the source of the stellar actinide-boost phenomenon to uncover whether this boost has an origin in key nuclear data or indicates some distinct \rp\ site.
In a very neutron-rich environment, such as the tidal ejecta from an NSM, we found that no single choice of nuclear input produces consistent radioactive decay ages for the very actinide-boost star, J0954$+$5246.
Instead, the actinide-to-lanthanide ratio is always overproduced compared to the observational abundance ratios, indicating the need for the actinides to be diluted by a lanthanide-rich, actinide-poor component of the \rp\ ejecta.
Such a component could be explained by the moderately neutron-rich accretion disk wind from an NSM remnant \cite{miller2019}.
The AD model builds such a possible ejecta distribution which dilutes the actinides of a tidal-only ejecta scenario with a lanthanide-rich wind component.
Borrowing values from both literature studies of NSM ejecta and estimates from the observed kilonova associated with GW170817, the AD model produces a better match to the observed Th/Eu abundance ratio of an actinide-boost star, especially by using the FRDM2012 mass model with consistent nuclear data.

Developing this model further, we used the ADM model to build empirical ejecta distributions describing how the material from an \rp\ event would be distributed in order to match actinide-boost abundances.
The results of this model indicate a significant, but non-dominant, component of the ejecta must have very low $Y_e$ in order to explain actinide-boost abundances.
However, from the smooth growth of the neutron-rich portion of the \rp\ ejecta in Figure~\ref{fig:adm}, there appears to be no distinct difference between actinide-boost cases and stars without an actinide-boost, indicating there is no need for a separate \rp\ progenitor type for actinide-boost stars.
Instead, the same type of \rp\ site can produce either an actinide-poor, actinide-normal, or actinide-boost case depending on the morphology of the ejecta.
For example, this can be accommodated in an NSM scenario if the merger event ejects more (actinide-boost) or less (actinide-poor) dynamical/tidal ejecta, which tend to be more neutron-rich than the wind ejecta.

\section*{References}
\bibliography{bibliography.bib}

\end{document}